# Attosecond Probing of Coherent Vibrational Dynamics in CBr$_4$

*Jen-Hao Ou,*[1] *Diptarka Hait,*[2,3] *Patrick Rupprecht,*[1,4] *John E. Beetar,*[1,†] *Todd J. Martínez,*[2,3] *and Stephen R. Leone*[1,4,5,*]

[1]Department of Chemistry, University of California, Berkeley, CA 94720, USA

[2]Department of Chemistry and The PULSE Institute, Stanford University, Stanford, California 94305, United States

[3]SLAC National Accelerator Laboratory, Menlo Park, California 94024, United States

[4]Chemical Sciences Division, Lawrence Berkeley National Laboratory, Berkeley, CA 94720, USA

[5]Department of Physics, University of California, Berkeley, CA 94720, USA

[†]Present address: NSF National eXtreme Ultrafast Science (NeXUS) Facility, Columbus, OH 43210, USA

E-mail: srl@berkeley.edu

**ABSTRACT**

A coherent vibrational wavepacket is launched and manipulated in the symmetric stretch ($a_1$) mode of $CBr_4$, by impulsive stimulated Raman scattering from non-resonant 400 nm laser pump pulses with various peak intensities on the order of tens of $10^{12}$ W/cm². Extreme ultraviolet (XUV) attosecond transient absorption spectroscopy (ATAS) records the wavepacket dynamics as temporal oscillations in XUV absorption energy at the bromine $M_{4,5}$ $3d_{3/2,5/2}$ edges around 70 eV. The results are augmented by nuclear time-dependent Schrödinger equation simulations. Slopes of the $(Br-3d_{3/2,5/2})^{-1}10a_1^*$ core-excited state potential energy surface (PES) along the $a_1$ mode are calculated to be -9.4 eV/Å from restricted open-shell Kohn-Sham calculations. Using analytical relations derived for the small-displacement limit with the calculated slopes of the core-excited state PES, a deeper insight into the vibrational dynamics is obtained by retrieving the experimental excursion amplitude of the vibrational wavepacket and the amount of population transferred to the vibrational first-excited state, as a function of pump-pulse peak intensity. Experimentally, the results show that XUV ATAS is capable of easily resolving oscillations in the XUV absorption energy on the order of few to tens of meV and tens of femtosecond time precision, limited only by the averaging times in the experimental scans. This corresponds to oscillations of C-Br bond length on the order of $10^{-4}$ to $10^{-3}$ Å. The results and the analytic relationships offer a clear physical picture, on multiple levels of understanding, for how the pump-pulse intensity controls the vibrational dynamics launched by non-resonant ISRS in the small-displacement limit.

**1. INTRODUCTION**

Preparing and controlling vibrational coherent superposition states (referred to as vibrational wavepackets) in molecules with light is a major theme in light-matter interactions.[1,2] This necessitates control over the creation, detection, and manipulation of the vibrational wavepacket.[1,2] In this work (Figure 1), such control is achieved in the carbon tetrabromide (CBr$_4$) molecule by attosecond transient absorption spectroscopy (ATAS).[3–5] We employ a 400 nm 26 fs pump laser pulse to generate a vibrational wavepacket in the symmetric stretch (a$_1$) mode of the electronic ground state of CBr$_4$ with non-resonant impulsive stimulated Raman scattering (ISRS).[6–8] The resulting vibrational wavepacket dynamics are subsequently monitored by time-resolved absorption of attosecond extreme ultraviolet (XUV) probe laser pulses at the Br M$_{4,5}$ 3d$_{3/2,5/2}$ edge around 70 eV, recorded as temporal oscillations in XUV absorption energy.

Similar vibrational dynamics in various molecules have been experimentally observed via transient absorption spectroscopy probed at XUV or soft X-ray regions.[5,9–28] Vibrational wavepackets were triggered by pump pulses mostly in the infrared (780, 800, 1200 or 1600 nm),[9–21,25–28] while some in visible[22] or UV[23,24] spectral regions. Among them, pump pulses with high (~$10^{14}$ W/cm$^2$) peak intensity[9,10,12–20,27,28] often lead to large vibrational amplitudes. In particular, relatively few works[26,27] systematically measured how the vibrational dynamics change with pump-pulse peak intensity. In this work, 400 nm pulses with lower peak intensities (8.1-20.2×$10^{12}$ W/cm$^2$) are employed, and the pump-pulse intensity dependence of vibrational dynamics is investigated in the small-displacement limit.[26]

To gain greater understanding, it is valuable to explicitly understand how the temporal oscillations in XUV absorption energy connect the shape of the corresponding core-excited state potential energy surface (PES) with aspects of the vibrational wavepackets, such as the excursion amplitude of the wavepacket and populations in the vibrational eigenstates. To clearly establish these connections, we derive analytical relations for the above quantities in the small-displacement limit with the harmonic approximation. We compute the slope of the Br 3d core-excited state PES along the symmetric stretch mode at the ground state equilibrium geometry from restricted open-shell Kohn-Sham (ROKS) calculations.[29–31] This slope is used to retrieve the excursion amplitude of the vibrational wavepacket (in terms of C-Br bond-length displacement) and the population transferred by ISRS to the vibrational first-excited state in the small-displacement limit, from the measured time-resolved XUV absorption spectrum.

Experimentally, the results show that XUV ATAS is capable of resolving small bond-length changes during vibrational wavepacket dynamics to the order of $10^{-4}$ Å with femtosecond temporal resolution. With analytical formulations, we offer a clear physical picture on multiple levels of understanding for how the pump-pulse peak intensity controls vibrational dynamics launched by non-resonant ISRS in the small-displacement limit.

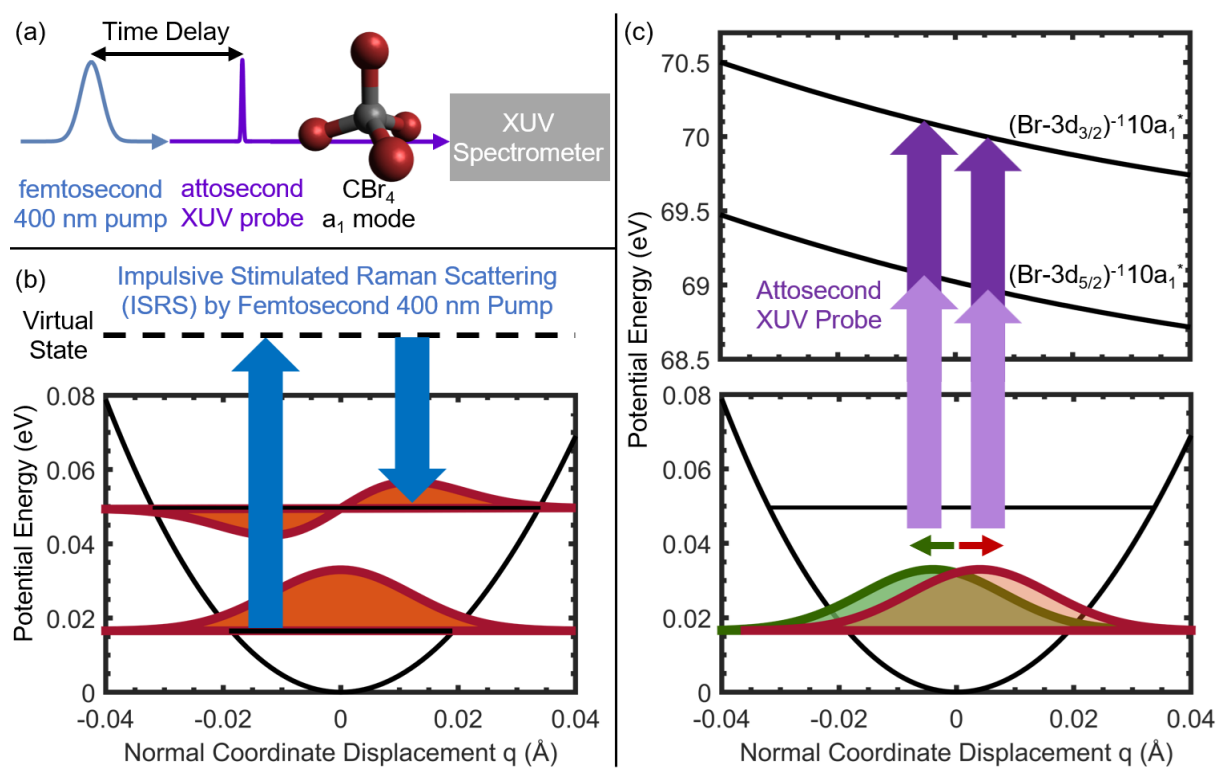

**Figure 1.** Experimental Scheme. (a) Attosecond transient absorption spectroscopy on the symmetric stretch ($a_1$) mode of $CBr_4$. (b) Vibrational eigenstates are coherently excited by non-resonant impulsive stimulated Raman scattering (ISRS) with a 400 nm 26 fs pump pulse, and a vibrational wavepacket is launched on the electronic ground state potential energy surface.

(c) Vibrational wavepacket dynamics are monitored by time-resolved absorption of attosecond XUV probe pulses from the electronic ground state to the $(Br-3d_{3/2,5/2})^{-1}10a_1^*$ core-excited states around 70 eV. The wavepackets drawn in green and red represent the motion shifting toward left and right, respectively.

## 2. METHODS

### 2.1 Experiment

The scheme of attosecond transient absorption spectroscopy on $CBr_4$, the 400 nm pump pulse, and the XUV probe pulse, are briefly outlined here and in Figure 1. The laser source is a Coherent Astrella USP (800 nm, 35 fs FWHM, 7 mJ, 1 kHz), and its output is split by a 70:30 beamsplitter into pump and probe arms.

#### 2.1.1 400 nm Pump Pulse

The pump arm starts with 70% of the Astrella output (800 nm, 35 fs FWHM, 4.9 mJ) and undergoes second harmonic generation (SHG) in a β-barium borate (BBO) crystal and the central wavelength becomes 400 nm. Then it is free-space focused through a Ne-filled (1.7 bar) tube (1.37 m long) to gently broaden the spectrum, and then compressed by chirped mirrors. The pulse duration of the main peak in the intensity envelope is measured to be 26±1 fs FWHM by homebuilt Self-Diffraction Frequency Resolved Optical Gating (SD-FROG) (Figure S6, Supporting Information).

#### 2.1.2 Few-Cycle Visible-NIR Pulse

The probe arm starts with 30% of the Astrella output (800 nm, 35 fs FWHM, 2.1 mJ) and is focused into a Ne-filled (2.4 bar) hollow core fiber (HCF) (1.5 m long, 400 µm inner diameter), where it undergoes supercontinuum generation, and then the pulses are compressed by chirped mirrors. After the chirped mirrors, the spectrum spans from around 500 to 900 nm. The dispersion is further fine tuned by a piece of 2 mm thick ADP crystal and a pair of thin glass wedges. The output is few-cycle (~ 4 fs) visible-NIR pulses that are used as the driving field for the high harmonic generation (HHG) into XUV region.

### 2.1.3 XUV Probe Pulse

The few-cycle visible-NIR pulses (driving field) are focused into a Ne gas cell (~80 mmHg, 300 µm hole diameter) and an XUV probe attosecond pulse train (~50-85 eV photon energy) is produced by high harmonic generation (HHG). The remaining 800 nm driving field is removed by an Al filter (0.1 µm thick).

### 2.1.4 Static and Time-Resolved (Transient) XUV Absorption Measurement

At the sample cell (4 mm long, 500 µm hole diameter), 400 nm pump and XUV probe beams recombine non-collinearly and interact with $CBr_4$ vapor. The sample reservoir, delivery line, and cell are heated to 100-120 °C to obtain enough vapor pressure from solid $CBr_4$. The time delay between the pump and probe pulses is controlled by changing the optical path length between the pump and probe beamlines with a translational stage. After a Zr filter (0.1 µm thick), the XUV beam is dispersed by a grating and detected on a XUV CCD camera.

Two types of XUV absorption spectra are collected experimentally. One is the static XUV absorbance $A(E_{XUV})$,

$$A(E_{XUV}) = -\log_{10} \frac{I_{XUV}(E_{XUV})}{I_{XUV}^{(0)}(E_{XUV})} \tag{1}$$

where $E_{XUV}$ is the photon energy of XUV probe pulses, and $I_{XUV}$ and $I_{XUV}^{(0)}$ are the XUV intensity with and without the $CBr_4$ molecule in the cell, respectively. The other is the time-resolved (transient) XUV differential absorption $\Delta A(E_{XUV}, \tau)$,

$$\Delta A(E_{XUV}, \tau) = -\log_{10} \frac{I_{XUV+pump}(E_{XUV}, \tau)}{I_{XUV}(E_{XUV}, \tau)} \tag{2}$$

where $\tau$ is the time delay between the pump and probe pulses, and $I_{\text{XUV+pump}}$ and $I_{\text{XUV}}$ are the XUV spectrum of CBr$_4$ with and without the pump pulse.

Temporal resolution is calibrated with a cross-correlation measurement of pump and probe pulses on the He 2s2p state. The fitted instrument response function is 26±6 fs and the time zero is -1±3 fs. The spectral axis is calibrated with a static XUV absorption spectrum of He 2snp (n ≥ 2) states. Details are in the Supporting Information.

### 2.1.5 Pump-Pulse-Intensity Dependent Measurement

The peak intensity of the 400 nm pump pulse is modified by cropping the beam with an iris, before the pump and probe pulses recombine. The peak intensity at the sample cell location is estimated to range from 8.1 to $20.2 \times 10^{12}$ W/cm$^2$. Detailed calibrations are given in the Supporting Information.

### 2.2 Calculation

Quantum chemical calculations are performed with the Q-Chem software package,[32] utilizing the SCAN0 functional[33] and the decontracted (i.e. all primitives are uncontracted) aug-cc-pVTZ basis set.[34–36] Scalar relativistic effects are included through the use of the spin-free exact two component one-electron (SF-X2C-1e) approach.[37–40] Local exchange-correlation integrals are calculated by quadrature over a radial grid with 250 points and an angular Lebedev grid with 974 points. The electronic ground state is modeled with the standard restricted Kohn-Sham procedure,[41] yielding an equilibrium bond distance of 1.925 Å (vs an experimental[42] value of of 1.942 Å) and a symmetric stretch frequency of 282 cm$^{-1}$ (vs 267 cm$^{-1}$ from experiment[43]).

The isotropic polarizability of the molecule is computed via finite differences of the energy with a five-point stencil (using electric field strengths of 0.005 and 0.01 a.u., see Supporting Information) over bond lengths ranging from 1.8246−2.046 Å (in increments of

0.001 Å). These polarizabilities are subsequently utilized to carry out nuclear time-dependent Schrödinger equation simulations of the ISRS process induced by the pump laser pulse and the resulting nuclear dynamics along the symmetric stretching mode, as described in detail in the Supporting Information.

The Br 3d excitation energies are computed with the ROKS approach,[29,30] which has been found to be quite effective at modeling core-level spectroscopies.[31,44] Excited state orbital optimization is carried out with the square gradient minimization algorithm.[45] Spin-orbit effects are subsequently incorporated in the manner described previously[31]. This quasi-degenerate perturbation-theory-based approach constructs a zeroth-order effective Hamiltonian of the spin-orbit free 3d excitation energies at a given geometry, to which a spin-orbit coupling operator $-J\,\vec{L}\cdot\vec{S}$ is added (where $\vec{S}$ is the electron spin operator and $\vec{L}$ the orbital angular momentum for d electrons). The coupling constant $J$ is assumed to be constant for all geometries and chosen to be 0.40 eV since this value reproduces the experimentally observed multiplet splitting of 1.03 eV between the $3d_{5/2}$ and $3d_{3/2}$ absorption energies. Core-level excitation energies are computed over bond lengths ranging from 1.9046−1.9446 Å, in increments of 0.001 Å, and are fitted to a straight line against the bond length (see Figure 3) in order to determine the slope of the core-excited PES vs bond elongation.

## 3. RESULTS AND DISCUSSION

**3.1 Detection of the Coherent Vibrational Dynamics: Static and Time-Resolved XUV Absorption Spectrum of CBr$_4$ at Br M$_{4,5}$ 3d$_{3/2,5/2}$ Edge**

Figure 2 (left) shows the static XUV absorption spectrum of CBr$_4$ from the electronic ground state to the (Br-3d$_{3/2,5/2}$)$^{-1}$10a$_1^*$ core-excited states (10a$_1^*$ is the lowest energy antibonding orbital[46,47]), and the spin-orbit doublet is resolved by fitting to a sum of two Voigt functions. The extracted peak energies are 70.932±0.011 and 69.903±0.008 eV for the Br M$_4$ 3d$_{3/2}$ and

$M_5$ 3$d_{5/2}$ edges, respectively (three decimal places are included here to compare with the very small (few to tens of meV) oscillation amplitude of XUV absorption energy measured and discussed below). The spin-orbit splitting is 1.03±0.01 eV (Table 1). They agree reasonably with previous experimental results[46] (70.5 and 69.6 eV). ROKS calculations yield slightly lower excitation energy values of 70.05 and 69.02 eV, indicating a somewhat larger difference between theory and the current experiment (~0.9 eV) than the ~0.3 eV deviations previously observed for the K- and L- edges of lighter elements.[31,44]

Next, we consider changes to the XUV absorption due to the vibrational wavepacket launched by 400 nm pump pulse (peak intensity 20.2×10$^{12}$ W/cm$^2$). As the vibrational wavepacket oscillates on the electronic ground state PES, the XUV absorption peak oscillates correspondingly with the vibrational period, as shown in the time-resolved XUV absorption spectrum in Figure 2 (right). The fitted periods are 123±1 fs for both the Br $M_4$ 3$d_{3/2}$ edge and $M_5$ 3$d_{5/2}$ edge, which agrees with the literature value of 126 fs for the symmetric stretch mode (267 cm$^{-1}$).[43] The oscillations in the peak position for the two edges are almost temporally in phase with a 0.1±0.4 rad phase difference. Oscillatory behaviors corresponding to the other normal modes of the molecule are not observed (despite such modes being Raman active) because there is no net displacement along such asymmetric modes, as has been observed and discussed in a previous work for $CCl_4$.[28]

If the launching mechanism of the vibrational wavepacket is a pure non-resonant ISRS from a Gaussian pump pulse centered at time zero, the expectation value of wavepacket position will be a sine function with a zero initial phase.[6,8] Nuclear TDSE simulations reproduce this expectation by retrieving around 0.01 π initial phase of wavepacket position for all intensities, using the temporal intensity envelope of the pump pulse retrieved from SD-FROG (Figure S6). Experimentally, over the range of intensities measured, the initial phase of XUV absorption energy varies from -0.1 to -0.6 π (Figure S12, Supporting Information), suggesting a possible deviation from the pure non-resonant ISRS. Potential causes may be the pre-pulse (Figure S6) or multiphoton absorption observed in the time-resolved XUV differential absorption spectrum (Figure S11).

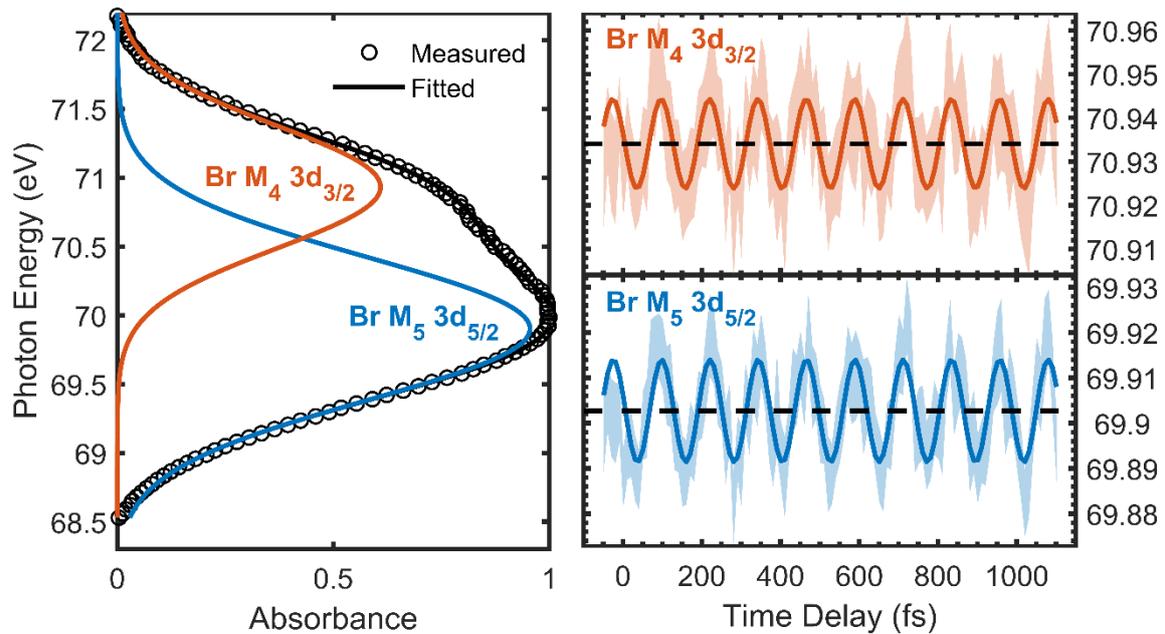

**Figure 2.** (Left) Static XUV absorption spectrum of CBr$_4$ at Br M$_{4,5}$ 3d$_{3/2,5/2}$ edges from the electronic ground state to the (Br-3d$_{3/2,5/2}$)$^{-1}$10a$_1^*$ core-excited states. The spin-orbit doublet is resolved by fitting to a sum of two Voigt functions. The extracted peak energies are 70.932±0.011 and 69.903±0.008 eV for the Br M$_4$ 3d$_{3/2}$ and M$_5$ 3d$_{5/2}$ edges respectively, with a 1.03±0.01 eV spin-orbit splitting resulting from the fit. (Right) Time-resolved XUV absorption spectrum of CBr$_4$ as a function of time delay between the 400 nm pump and XUV probe pulses. The peak intensity of pump pulse is 20.2×10$^{12}$ W/cm$^2$. Color shade: extracted peak energy with error bars at each time delay. Solid line: a cosine wave fitted to the extracted peak energies. The fitted period is 123±1 fs, amplitude 10.2±1.4 meV, center 70.934±0.001 eV (black dashed line) for the M$_4$ 3d$_{3/2}$ edge, while the M$_5$ 3d$_{5/2}$ edge yields 123±1 fs, 11.3±1.4 meV, and 69.903±0.001 eV (black dashed line). The oscillations for the two edges are almost temporally in phase with a 0.1±0.4 rad phase difference.

**Table 1.** Excitation energies (eV) of (Br-3d$_{3/2,5/2}$)$^{-1}$10a$_1{}^*$ core-excited states in CBr$_4$.

| Core-Excited State | Experiment (This work) | ROKS Calculations (This work) | Previous Experiment[46] |
|---|---|---|---|
| (Br-3d$_{3/2}$)$^{-1}$10a$_1{}^*$ | 70.932±0.011 | 70.05 | 70.5 |
| (Br-3d$_{5/2}$)$^{-1}$10a$_1{}^*$ | 69.903±0.008 | 69.02 | 69.6 |
| Spin-Orbit Splitting | 1.03±0.01 | 1.03 | 0.9 |

**3.2 Information Encoded in the Time-Resolved XUV Absorption Spectrum: Oscillation Amplitude of XUV Absorption Energy, Excursion Amplitude of Vibrational Wavepacket, and Slope of Core-Excited State Potential Energy Surface**

When the vibrational wavepacket motion results in a normal mode displacement $q$ on the electronic ground state PES $E_g(q)$, the XUV absorption energy $E_{XUV}(q)$ from the electronic ground state to the core-excited state PES $E_{ce}(q)$ is $E_{XUV}(q) = E_{ce}(q) - E_g(q)$. If the pump-pulse intensity is not too strong to induce a large displacement $q$, the harmonic approximation holds and the XUV absorption energy can be approximated by a Taylor series in $q$

$$E_{XUV}(q) \approx \left[ E_{ce}(0) - E_g(0) \right] + \left[ \left( \frac{dE_{ce}}{dq} \right)_0 - \cancel{\left( \frac{dE_g}{dq} \right)_0} \right] q$$
$$= E_{XUV}(0) + \left( \frac{dE_{ce}}{dq} \right)_0 q \qquad (3)$$

where the slope of the electronic ground state PES $E_g(q)$ vanishes at the ground state equilibrium position $q = 0$, while the slope of the electronic core-excited state PES $E_{ce}(q)$ will generally be nonzero at the Frank-Condon region, as the equilibrium position of the core-excited PES is quite unlikely to coincide with that of the ground state.

As the vibrational wavepacket travels between the outer turning point $q_{ot}$ to the inner turning point $q_{it}$ on the electronic ground state PES, the range of peak XUV absorption energies that can be accessed is

$$\left| E_{XUV}(q_{ot}) - E_{XUV}(q_{it}) \right| = \left| \left( \frac{dE_{ce}}{dq} \right)_0 \right| (q_{ot} - q_{it}) \tag{4}$$

where $\Delta E_{XUV} \equiv \left| E_{XUV}(q_{ot}) - E_{XUV}(q_{it}) \right| / 2$ is defined as the *oscillation amplitude of the XUV absorption energy*, and $q_{amp} \equiv (q_{ot} - q_{it})/2$, the *excursion amplitude of the vibrational wavepacket*, or physically the maximum displacement from equilibrium. Therefore,

$$\boxed{\Delta E_{XUV} = \left| \left( \frac{dE_{ce}}{dq} \right)_0 \right| q_{amp}} \tag{5}$$

and $q_{amp}$ can thus be retrieved from $\Delta E_{XUV}$ in the experimentally measured time-resolved XUV absorption spectrum in Figure 2 (right), if the slope of the core-excited state PES along the normal mode at the electronic ground state equilibrium geometry is known.

In this work, the slope of the (Br-3d$_{3/2,5/2}$)$^{-1}$10a$_1^*$ core-excited states PES is calculated from ROKS to be around -9.4 eV/Å along the symmetric stretch mode (by fitting core-excitation energies to a straight line, as shown in Figure 3). In other words, this slope suggests that the Br-3d$_{3/2,5/2}$→10a$_1^*$ excitation energy decreases by 9.4 eV as the C-Br bond length increases by 1 Å. This is consistent with the antibonding character of the 10a$_1^*$ orbital, as the energy of this state should significantly decrease with bond elongation. This 9.4 eV/Å value for the slope is in the same order of magnitude (~10 eV/Å) as previous estimates for core-to-antibonding excitations in $CCl_4$[28] and $SF_6$[26,27] (typical range 5 – 30 eV/Å depending on which atom).

The ROKS results and the resulting linear fits are shown in Figure 3, which indicates that Equation 3 holds quite well for bond stretches as high as ±0.05 Å relative to equilibrium and higher order effects do not appear to be relevant in this regime. The ~10 meV or smaller experimental oscillation amplitudes for XUV absorption energies (Figure 2) and the computed slope -9.4 eV/Å collectively indicate that the bond stretches remain on the order of $10^{-4}$ Å, and the linear behavior reported in Equation 3 therefore is sufficient for analyzing the experimental results.

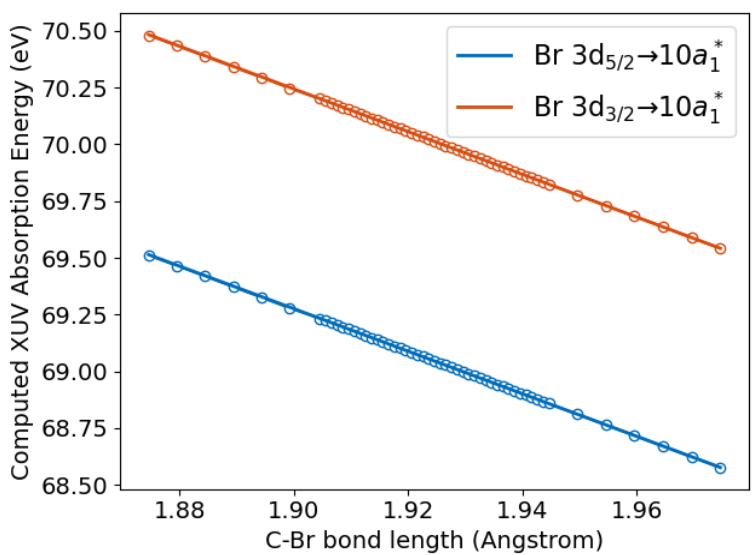

**Figure 3.** Computed XUV excitation energies from ROKS for various symmetric C-Br bond stretches (open circles) around the equilibrium geometry (1.925 Å). The linear fits obtained from fitting to ROKS results for 41 equally spaced bond lengths between 1.9046-1.9446 Å are also shown (solid lines), which align quite well with ROKS results for twelve bond lengths calculated outside the fit interval (six equally spaced points each between 1.8746-1.8996 Å and 1.9496-1.9746 Å). The slopes are about -9.4 eV/Å for the Br-$3d_{3/2,5/2}$→$10a_1^*$ excitations.

**3.3 Relative Slope between Multiple Core-Excited State PES and Insight into Property of Participating Orbitals**

If multiple electronic core-excited states simultaneously probe the vibrational wavepacket on the electronic ground state, the *relative slope* between two or more core-excited state PES[26–28] can be determined experimentally from the ratio of measured oscillation amplitude of XUV absorption energy $\Delta E_{\text{XUV}}$ as

$$\Delta E_{\text{XUV}}^{\text{ce2}} / \Delta E_{\text{XUV}}^{\text{ce1}} = \left| \left( \frac{dE_{\text{ce2}}}{dq} \right)_0 \right| / \left| \left( \frac{dE_{\text{ce1}}}{dq} \right)_0 \right| \tag{6}$$

because here the excursion amplitude of the vibrational wavepacket $q_{\text{amp}}$ is the same. In this work, the relative slope between the two $(\text{Br-3d}_{3/2,5/2})^{-1}10a_1^*$ core-excited states PES is determined experimentally to be 1.1±0.13 as the ratio of measured oscillation amplitude of XUV absorption energy $\Delta E_{\text{XUV}}$ (Br-3d$_{5/2}$)/ $\Delta E_{\text{XUV}}$ (Br-3d$_{3/2}$) (Table S4, Supporting Information). As a result, the shape of the two core-excited state PESs are quite similar around the equilibrium geometry of the electronic ground state.

Such relative slopes not only offer experimental information on the shape of core-excited state PES, but also shed light on the nature of participating orbitals. For the discussion below, a core-excited state is approximated as a singly-excited electronic configuration, where one electron is excited from an initial core orbital to a final orbital that is higher in energy.

One scenario compares cases when the electron is excited from the same core orbital to different final orbitals. The relative slopes of the corresponding core-excited state PESs belonging to the same absorption edge gives a hint about the bonding properties of the final orbitals. For instance, if the final orbital is non-bonding, then the resulting core-excited state will be generally less sensitive to changes in the molecular geometry and its PES will be relatively flat. As a result, its slope will be smaller than the slope of other core-excited states PES where the final orbital is bonding or anti-bonding. For instance, in the SF$_6$ molecule[26,27], an electron can be excited from S-2p core orbital to three different final orbitals $a_{1g}$, $t_{2g}$ and $e_g$.

Because the $a_{1g}$ and $e_g$ orbitals are anti-bonding while the $t_{2g}$ orbital is non-bonding, the slope of (S-2p)$^{-1}a_{1g}$ and (S-2p)$^{-1}e_g$ core-excited states is found to be larger than that of (S-2p)$^{-1}t_{2g}$ state.

The other scenario is when the electron is excited from different core orbitals to the same final orbital, or the corresponding core-excited states originating from different absorption edges. This offers a potential way to compare different core orbitals on the same atom, or core orbitals on different atoms in a molecule. In this work, the electron is excited to the same $10a_1^*$ anti-bonding orbital, but from different Br $3d_{3/2}$ or Br $3d_{5/2}$ core orbitals, corresponding to the Br $M_4$ or $M_5$ absorption edges. The ratio of slopes of these two core-excited states is close to unity as mentioned above, indicating that the both Br $3d_{3/2}$ or Br $3d_{5/2}$ core orbitals respond similarly to the bond-length change. This is reasonable as these two core orbitals come from the same set of Br 3d orbitals, split by spin-orbit coupling. On the other hand, a previous study on CCl$_4$ molecule[28] found the slope for the (Cl-2p$_{3/2}$)$^{-1}8t_2^*$ core-excited state to be much larger than that of the (C-1s)$^{-1}8t_2^*$ core-excited state, despite the same final orbital ($8t_2^*$) for both excitations.

### 3.4 Connection between the Excursion Amplitude of Vibrational Wavepacket and Vibrational Eigenstate Populations, in the Small-Displacement Limit

In the small-displacement limit, only the vibrational ground $\psi_0$ and first-excited $\psi_1$ states of a harmonic oscillator need to be considered.[26] Long after the pump pulse, the coherent vibrational wavepacket $\Psi(q, t)$ can be approximated by the following linear superposition,

$$\Psi(q,t) = c_0 \psi_0(q) e^{-iE_0 t/\hbar} + c_1 \psi_1(q) e^{-iE_1 t/\hbar} \qquad (7)$$

where $q$ is the normal mode displacement (position), and $E_{0,1}$ is the vibrational energy of the vibrational ground and first-excited states, respectively. The expectation value of position $\langle q(t) \rangle$ of this vibrational wavepacket is:

$$\begin{aligned}
&\langle q(t) \rangle \\
&= \langle \Psi(q,t)|q|\Psi(q,t) \rangle \\
&= p_0 \langle \psi_0|q|\psi_0 \rangle \\
&\quad + p_1 \langle \psi_1|q|\psi_1 \rangle \\
&\quad + 2\sqrt{p_0 p_1} \langle \psi_0|q|\psi_1 \rangle \cos[(E_1-E_0)t/\hbar + \phi]
\end{aligned} \quad (8)$$

where $p_{0,1} = |c_{0,1}|^2$ is the population of the vibrational ground and first-excited states, respectively, and $\phi$ is the initial phase between the two coefficients $c_{0,1}$. (If the launching mechanism of the coherent vibrational wavepacket is non-resonant ISRS from a Gaussian pump pulse centered at time zero, the expectation value of position $\langle q(t) \rangle$ will be a sine function with a zero initial phase,[6,8] or equivalently a cosine function with a -π/2 initial phase in Equation 8.)

Next, because $\langle \psi_0|q|\psi_0 \rangle = \langle \psi_1|q|\psi_1 \rangle = 0$ and $\langle \psi_0|q|\psi_1 \rangle = \sqrt{\frac{\hbar}{2\mu\omega_0}}$ for a harmonic oscillator (where $\mu$ is the mass associated with the normal mode, which is four times the mass of Br, and $\omega_0 = (E_1 - E_0)/\hbar$ is the fundamental vibrational frequency of the normal mode), the position expectation value simplifies to:

$$\begin{aligned}
\langle q(t) \rangle &= \left( 2\sqrt{p_0 p_1} \sqrt{\frac{\hbar}{2\mu\omega_0}} \right) \cos(\omega_0 t + \phi) \\
&= \left( 2\sqrt{(1-p_1)p_1} \sqrt{\frac{\hbar}{2\mu\omega_0}} \right) \cos(\omega_0 t + \phi) \\
&= q_{amp} \cos(\omega_0 t + \phi)
\end{aligned} \quad (9)$$

as $p_1 + p_0 = 1$. This shows that the excursion amplitude of the vibrational wavepacket is

$$q_{amp} = 2\sqrt{(1-p_1)p_1} \sqrt{\frac{\hbar}{2\mu\omega_0}} \quad (10)$$

Furthermore, when $p_1 \ll 1$ (as expected in the small-displacement limit),

$$\boxed{q_{amp} \approx 2\sqrt{p_1}\sqrt{\frac{\hbar}{2\mu\omega_0}}} \tag{11}$$

so $q_{amp}$ is proportional to $\sqrt{p_1}$. Therefore, if the excursion amplitude of the vibrational wavepacket is known, the population in the vibrational first-excited state can be determined from this relation.

The validity of the above formulation is also verified against numerical simulations of the nuclear time-dependent Schrödinger equation (TDSE) along the symmetric stretch mode alone. As shown in Figure 4, the calculated population of the vibrational ground state is close to one (on the order of 0.99), and the calculated population of the vibrational first-excited state is indeed small (around the order of $10^{-3}$) for the pump pulse intensities used in this work. As a result, the pump light-matter interaction studied in this work is clearly in the small-displacement limit.

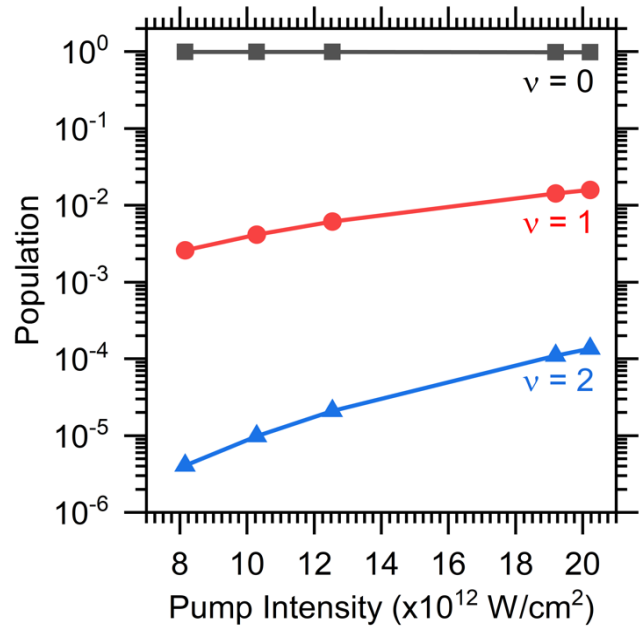

**Figure 4.** Calculated populations of the vibrational eigenstates ($v = 0, 1, 2$) of the symmetric stretch $a_1$ mode from nuclear TDSE simulations, as a function of the experimental 400 nm pump-pulse intensity. The population of the vibrational first-excited state remains small (around the order of $10^{-3}$). This shows that the pump light-matter interaction is in the small-

displacement limit for the range of pump-pulse intensities used in this work. The population in the second vibrational excited state is even smaller, indicating that the two-state approximation used for the small-displacement limit is reasonable.

**3.5 Physical Basis for the Phenomenological Equations in the Literature to Fit the Oscillations in Time-Resolved XUV Absorption Spectrum due to Coherent Vibrational Dynamics**

Based on the formulation above, a physical basis can be provided for the phenomenological equations that have often been used in the literature[9,12,14,23,24,27] to fit the oscillations in time-resolved XUV absorption spectrum due to coherent vibrational dynamics. If only one vibrational mode is involved, the time-resolved XUV absorption energy is often described empirically as

$$E_{XUV}(t) = E_{XUV}^0 + \Delta E_{XUV}\cos(\omega_0 t + \phi) \qquad (12)$$

where $E_{XUV}^0$ is the XUV absorption energy at the equilibrium geometry of electronic ground state, and $\Delta E_{XUV}$ is the oscillation amplitude of XUV absorption energy. According to Equation 5,

$$\Delta E_{XUV} = \left|\left(\frac{dE_{ce}}{dq}\right)_0\right| q_{amp} \qquad (5)$$

it further reveals that oscillation amplitude of XUV absorption energy is proportional to the slope of core-excited state PES along the normal mode at the electronic ground state equilibrium geometry and the excursion amplitude of vibrational wavepacket, within the harmonic approximation. Larger amplitudes thus can result from steeper slopes of the core-excited PES (which can be attained through excitation to an antibonding level, as in this work), or from higher pump-pulse peak intensity to access greater excursion amplitudes of vibrational wavepacket $q_{amp}$, by adding a greater fraction of the vibrational first-excited state in the small-displacement limit.

**3.6 Controlling the Motion of Vibrational Wavepacket with Different Pump-Pulse Peak Intensities in the Small-Displacement Regime**

The dynamics of the vibrational wavepackets are initiated with five different pump-pulse peak intensities ranging from 8.1 to $20.2 \times 10^{12}$ W/cm$^2$ (detailed calibrations provided in Table S1). Figure 5(a) shows that the experimentally measured oscillation amplitudes of XUV absorption energy increase approximately linearly with pump-pulse intensity, which is also observed for the SF$_6$ molecule[26,27].

In Figure 5(b), the excursion amplitudes of the vibrational wavepacket $q_{amp}$ are retrieved from the corresponding oscillation amplitudes of the XUV absorption energy $\Delta E_{XUV}$ by Equation 5, given the -9.4 eV/Å slope of the (Br-3d$_{3/2,5/2}$)$^{-1}$10a$_1{}^*$ core-excited state PES from ROKS calculations. In Figure 5(c), the square root of population in the vibrational first-excited state $\sqrt{p_1}$ is further estimated from the excursion amplitude of the vibrational wavepacket $q_{amp}$ via Equation 11. All three quantities plotted in Figure 5 are found to increase approximately linearly with the pump-pulse intensity for both the Br M$_4$ 3d$_{3/2}$ and M$_5$ 3d$_{5/2}$ edges.

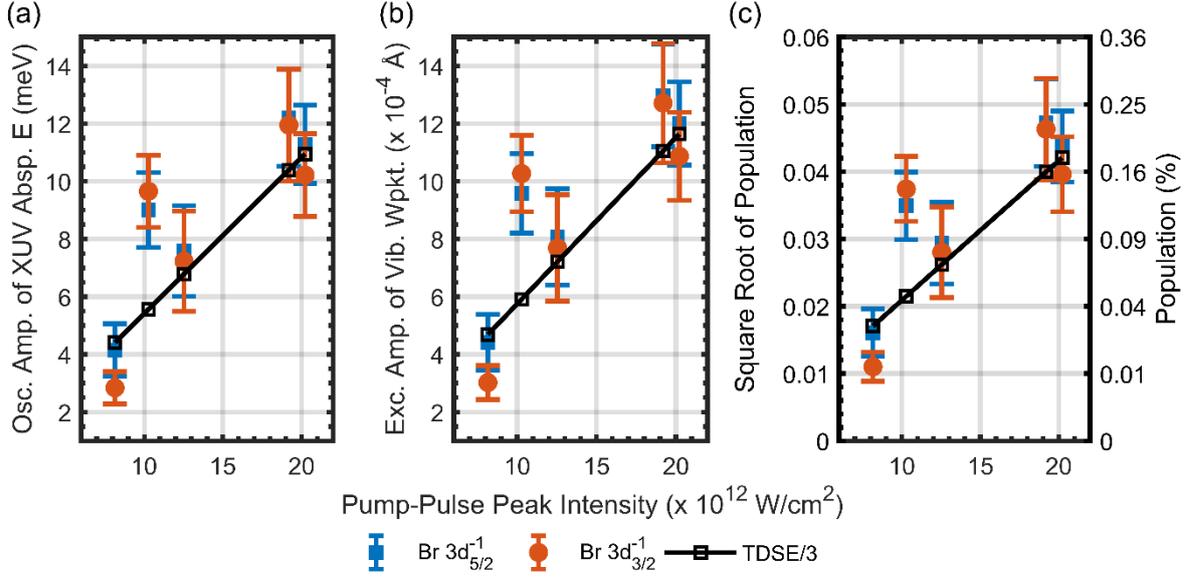

**Figure 5.** Pump-pulse peak-intensity dependence of (a) experimentally measured oscillation amplitude of the XUV absorption energy $\Delta E_{XUV}$, (b) excursion amplitude of the vibrational wavepacket $q_{amp}$ (retrieved from (a) by Equation 5, given the -9.4 eV/Å slope of the (Br-3d$_{3/2,5/2}$)$^{-1}$10a$_1^*$ core-excited state PES from ROKS calculations), and (c) square root of the population transferred to the vibrational first-excited state $\sqrt{p_1}$ (estimated from (b) by Equation 11). All three quantities increase approximately linearly with the pump-pulse intensity for both the Br M$_4$ 3d$_{3/2}$ (orange circle) and M$_5$ 3d$_{5/2}$ (blue square) edges. Estimates of all three quantities from nuclear TDSE simulations (black open square) for the ISRS process are also provided (scaled by 1/3, see text for details), and show clear linear relationships with the pump-pulse intensities used in this work. The outlier at $10.3\times10^{12}$ W/cm$^2$ may result from the fluctuation in average power measurement, as the average power is close to the lower limit of the working range for the power meter head (Coherent PowerMax PM10).

The experimentally measured and retrieved values are also compared with nuclear TDSE simulations. The theoretical population of the vibrational first-excited state $p_1$ and excursion amplitude of the vibrational wavepacket $q_{amp}$ come from numerically solving the nuclear TDSE for the non-resonant ISRS process, and the theoretical oscillation amplitude of XUV absorption energy $\Delta E_{\text{XUV}}$ is estimated from Equation 5. The TDSE simulation results have to be scaled by 1/3 to match the experiment. A few possible reasons for this large discrepancy between experiment and theory are considered below. We note that the TDSE simulations account for rotational averaging effects on the interaction between the molecule and the pump pulse field through the use of isotropic polarizability.

From an experimental perspective, the differences with theory may arise from a lower effective pump-pulse peak intensity at the sample location than the estimation. A detailed calibration of the experimental peak intensity for the pump pulse is given in the Supporting Information, which is taken as our best estimation. However, it is uncertain whether the pump and probe pulses spatially overlap right at the beam center of each other. If the spatial overlap is off-center, the effective pump-pulse peak intensity within the probe beam area will become smaller. Before the measurements on CBr$_4$ were made, the spatial overlap was confirmed by overlapping the recombined pump beam and the HHG driving-field beam on a beam profiler at a location with equivalent distance to the sample cell, as the XUV probe beam most likely follows the same path of the HHG driving-field beam. However, if there is an angular mismatch, the overlap could be diminished and therefore be a cause of the discrepancy. On the other hand, the outlier at $10.3 \times 10^{12}$ W/cm$^2$ may result from the fluctuations in average power measurement, as the average power is close to the lower limit of the working range for the power meter head (Coherent PowerMax PM10).

Potential computational origins of this discrepancy are errors in the Br 3d core-excited PES slope value of -9.4 eV/Å, or inadequacies in modeling the non-resonant ISRS process through TDSE (such as from errors in the molecular polarizabilities). We consider these factors to be somewhat less likely as the slope is quite consistent with previous results for other molecules, and additional effort has been invested into benchmarking molecular polarizabilities used for the TDSE simulations (see Supporting Information).

## 4. CONCLUSIONS

In this work, a coherent vibrational wavepacket is launched in the symmetric stretching ($a_1$) mode of electronic ground state $CBr_4$ molecule by non-resonant impulsive stimulated Raman scattering (ISRS) with 400 nm pump pulses. Dynamics of this vibrational wavepacket are studied with time-resolved absorption of attosecond XUV pulses at the bromine $M_{4,5}$ $3d_{3/2,5/2}$ edges around 70 eV, with excitations to the lowest energy antibonding ($10a_1^*$) orbital. The measured XUV absorption energies oscillate in time with the period of the symmetric stretching mode (126 fs), and the oscillation amplitude of XUV absorption energy equals the product of the excursion amplitude of the vibrational wavepacket and the slope of the Br (Br-$3d_{3/2,5/2})^{-1}10a_1^*$ core-excited state potential energy surface along this mode. Since the measured oscillation amplitude of XUV absorption energy is on the order of few to tens of meV and the slope is -9.4 eV/Å from ROKS calculations, the excursion amplitude of the vibrational wavepacket (C-Br bond-length displacement) is retrieved to be on the order of $10^{-4}$ to $10^{-3}$ Å. Harmonic approximation therefore holds quite well in this small-displacement limit.

Experiments with various pump-pulse peak intensities (8.1-20.2×$10^{12}$ W/cm$^2$) show an approximately linear relationship between the intensity and the measured oscillation amplitude of XUV absorption energy, and between the intensity and the retrieved excursion amplitude of vibrational wavepacket as well. Furthermore, the excursion amplitude of vibrational wavepacket is proportional to the square root of the population in the vibrational first-excited state in the small-displacement limit. This relation enables an indirect estimate of this population from the measured oscillation amplitude of XUV absorption energy. These linear relationships are supported by numerical simulations of the nuclear time-dependent Schrodinger equation along the symmetric stretching mode for the non-resonant ISRS.

The present analysis and approach to extract information encoded in time-resolved XUV absorption spectra is general and can be readily applied to ISRS combined with time-resolved absorption spectroscopy probed in other spectral regions. Experimentally, the results show that XUV attosecond transient absorption spectroscopy is capable of resolving the small bond-length changes during vibration with extraordinary $10^{-4}$ to $10^{-3}$ Angstrom and tens of femtosecond precision.[26] Moreover, the analytical expressions provide a concrete and thorough understanding of how the pump-pulse intensity controls vibrational dynamics in the small-displacement limit.

**ACKNOWLEDGMENT**

J.H.O., J.E.B., and S.R.L. thank the National Science Foundation under grant CHE-2243756 for support of the experimental work. J.H.O. thanks the Government Scholarship to Study Abroad, Ministry of Education, Taiwan for supplemental support, and Stefan Schippers and Christian Schroeder for discussions on fitting Fano-Gaussian lineshapes. D.H. is a Stanford Science Fellow. is. P.R. acknowledges funding by the Alexander von Humboldt Foundation (Feodor-Lynen Fellowship) and support of the AMOS program of the Office of Science of the U.S. Department of Energy under Contract No. DE-AC02-05CH11231. T.J.M and D.H. were supported by the AMOS program of the Office of Science of the U. S. Department of Energy. This research used resources of the National Energy Research Scientific Computing Center, a DOE Office of Science User Facility supported by the Office of Science of the U.S. Department of Energy under Contract No. DE-AC02-05CH11231 using NERSC award BES-ERCAP0027716.

# SUPPORTING INFORMATION AVAILABLE

1. Calibration of Pump Pulse (Beam diameter, average power, peak intensity, and Self-Diffraction Frequency Resolved Optical Gating (SD-FROG) measurement)

2. Calibration of Time Zero and Temporal Resolution for Time-Resolved XUV Absorption Measurements

3. Calibration of Spectral Axis

4. Measured Relative Slope of Br-3d Core-Excited State Potential Energy Surface (PES)

5. Comment about Initial Phase of Wavepacket Trajectory

6. Nuclear Time-Dependent Schrodinger Equation Simulations

7. Comment about Polarizability Calculations

# Supporting Information for "Attosecond Probing of Coherent Vibrational Dynamics in CBr$_4$"

*Jen-Hao Ou,*[1] *Diptarka Hait,*[2,3] *Patrick Rupprecht,*[1,4] *John E. Beetar,*[1,†] *Todd J. Martínez,*[2,3] *and*

*Stephen R. Leone*[1,4,5,*]

[1]Department of Chemistry, University of California, Berkeley, CA 94720, USA

[2]Department of Chemistry and The PULSE Institute, Stanford University, Stanford, California 94305, United States

[3]SLAC National Accelerator Laboratory, Menlo Park, California 94024, United States

[4]Chemical Sciences Division, Lawrence Berkeley National Laboratory, Berkeley, CA 94720, USA

[5]Department of Physics, University of California, Berkeley, CA 94720, USA

[†]Present address: NSF National eXtreme Ultrafast Science (NeXUS) Facility, Columbus, OH 43210, USA

E-mail: srl@berkeley.edu

# 1. Calibration of Pump Pulse

In this work, the peak intensity of pump pulse is modified by cropping the beam with an iris before the pump and probe pulses recombine. Detailed measurements of pump-pulse properties are given below.

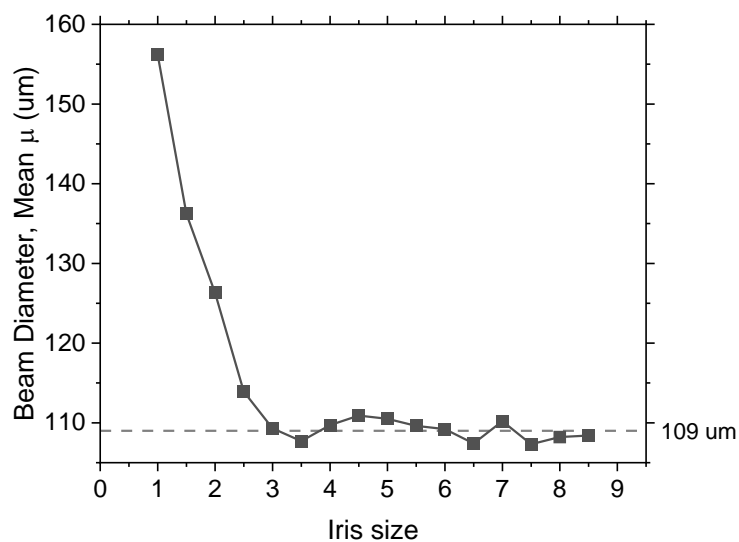

**Figure S1.** Beam diameter (mean µ) of pump pulse vs iris size. The iris size is read by the numbered marks on the iris. Measured with DataRay beam profiler WinCamD-UHR, 4xsigma mode. Picked off to the location at the same distance to the sample location. The beam diameter (mean µ) remains around 109 µm and does not change significantly when the iris size decreases from full to 3.0. Only when iris size is smaller than 3.0, the beam starts to get significantly cropped and the beam diameter increases appreciably. The averaged value of beam diameters from size 3.0 to full is 109 µm with 1 µm standard deviation.

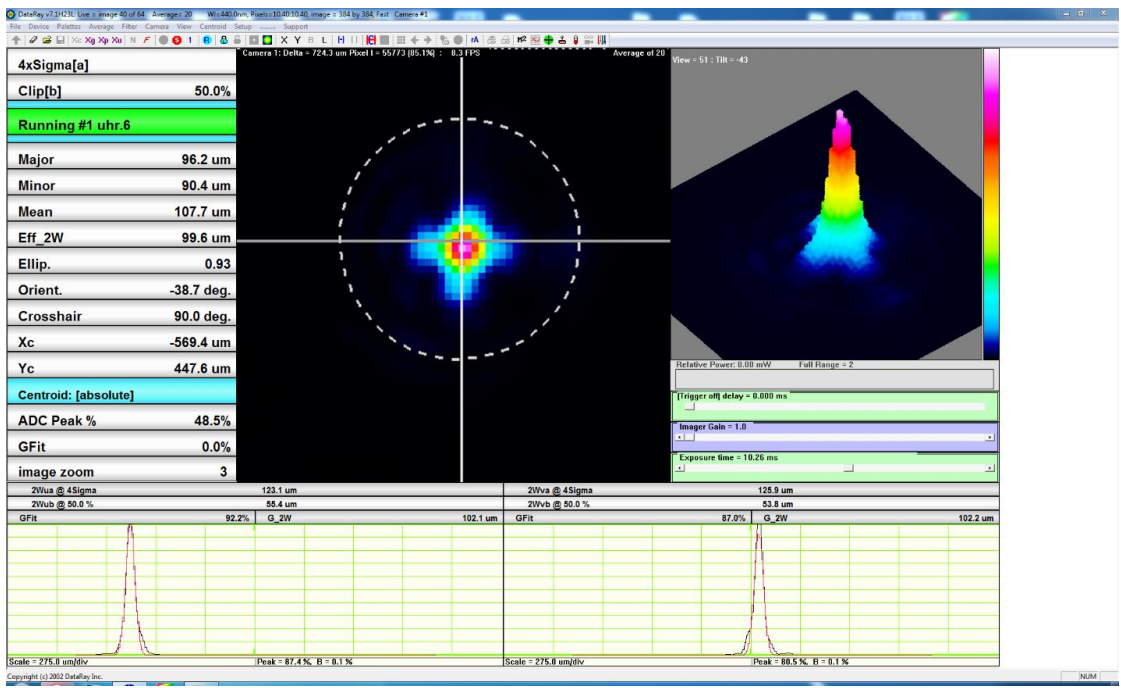

**Figure S2.** Beam shape with iris size full measured by DataRay beam profiler WinCamD-UHR. The mean is 107.7 μm.

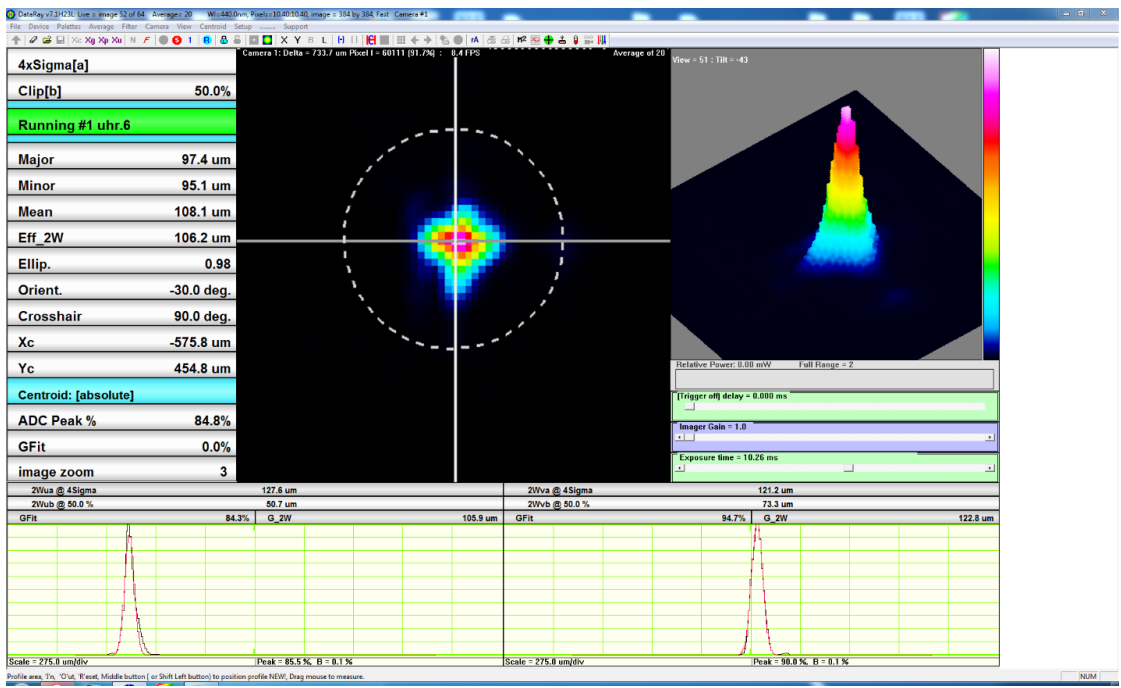

**Figure S3.** Beam shape with iris size 3.5 measured by DataRay beam profiler WinCamD-UHR. The mean is 108.1 μm.

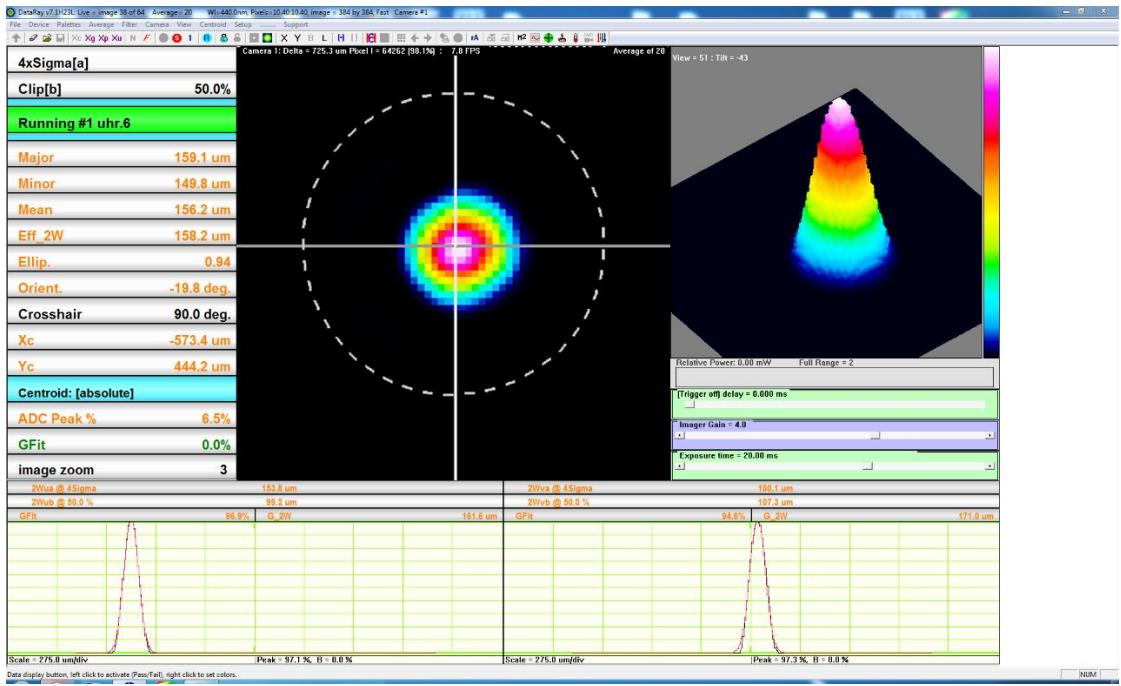

**Figure S4.** Beam shape with iris size 1.0 measured by DataRay beam profiler WinCamD-UHR. The mean is 156.2 μm.

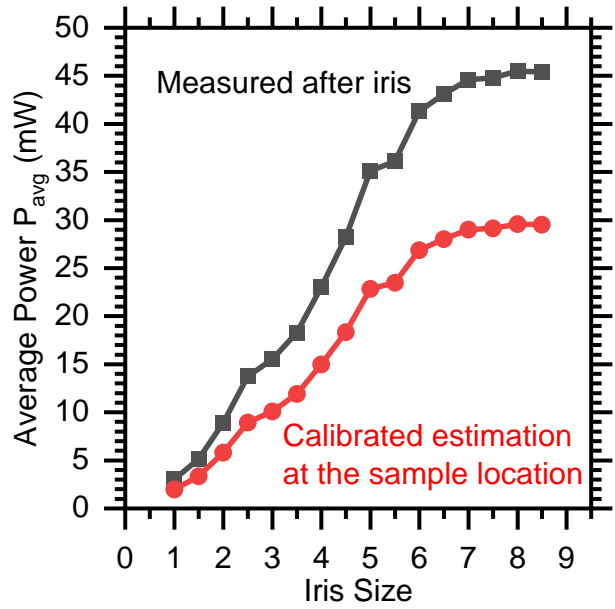

**Figure S5.** Average power of pump pulse vs iris size. The iris size is read by the numbered marks on the iris. The average power at the sample location (red circle) is estimated as 65% of that measured right after the iris (black square), according to a previous power calibration. There are 5 mirrors and 1 thin glass window between the iris and the sample cell.

**Table S1.** Peak intensity of pump pulse vs iris size.

| Scan # | Iris size[a] | Average Power (mW)[b] | Beam Diameter, Mean μ (μm)[c] | Pulse Duration (fs)[d] | Pulse Energy (μJ)[e] | Peak Intensity (×$10^{12}$ W/cm²)[f] | Peak Intensity (×$10^{-4}$ atomic unit)[g] |
|---|---|---|---|---|---|---|---|
| 2 | 3.5 | 11.9 | 109±1 | 25.8±0.6 | 11.9 | 8.15 | 2.32 |
| 3 | 4 | 15.0 | 109±1 | 25.8±0.6 | 15.0 | 10.3 | 2.93 |
| 5 | 4.5 | 18.3 | 109±1 | 25.8±0.6 | 18.3 | 12.5 | 3.57 |
| 4 | 6.5 | 28.0 | 109±1 | 25.8±0.6 | 28.0 | 19.2 | 5.47 |
| 1 | full | 29.5 | 109±1 | 25.8±0.6 | 29.5 | 20.2 | 5.76 |

[a] Iris size is read by the numbered marks on the iris.

[b] At the sample cell location. Estimated according to previous power calibration (Figure S5)

[c] Measured with DataRay beam profiler WinCamD-UHR, 4xsigma mode. Details in Figure S1.

[d] Measured by SD-FROG (Figure S6(d))

[e] Average power divided by 1 kHz repetition rate

[f] Assume Gaussian spatial profile and FROG-retrieved temporal profile (Figure S6)

[g] Peak intensity in W/cm² divided by the conversion factor 3.50944758×$10^{16}$ from the literature[1].

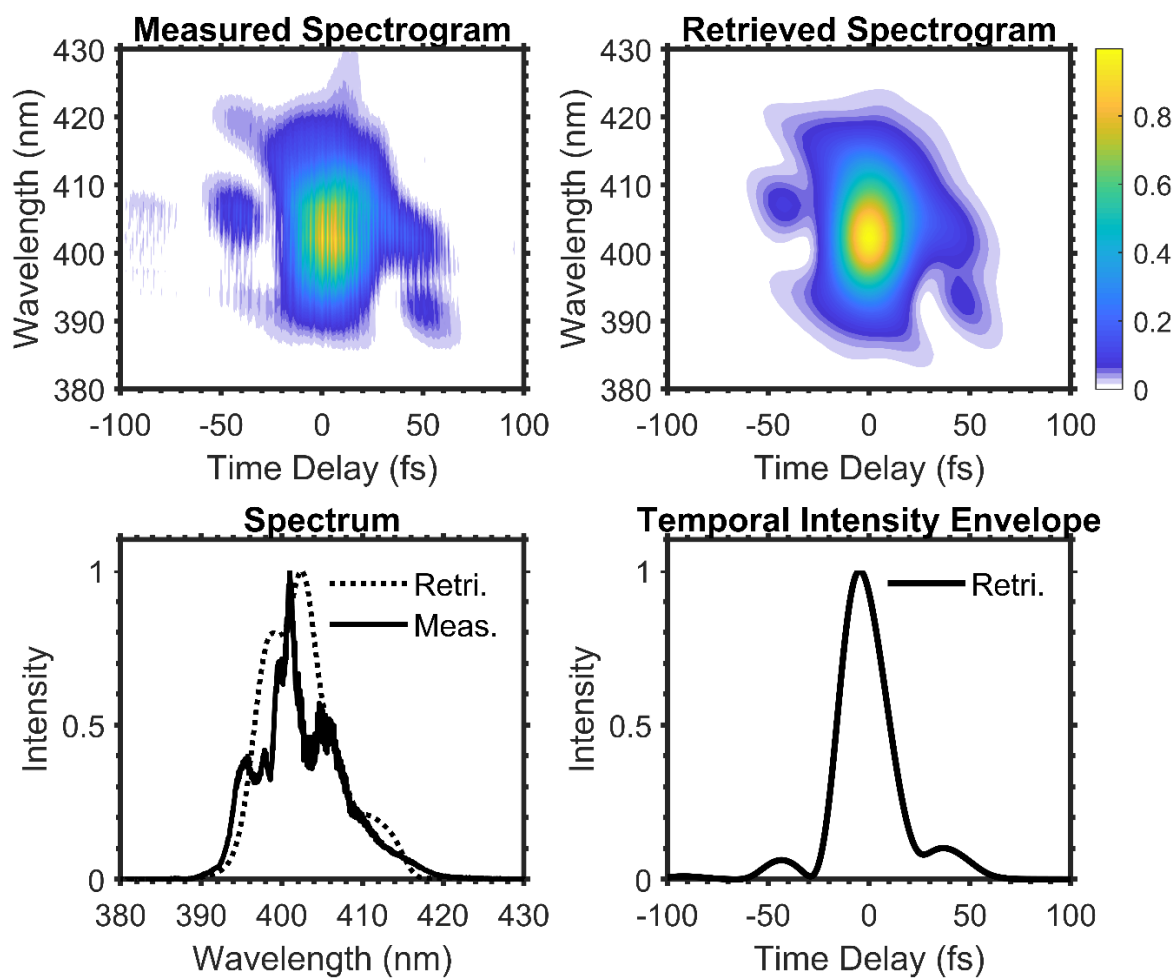

**Figure S6.** Self-Diffraction Frequency Resolved Optical Gating (SD-FROG) measurement for 400 nm pump pulse (a) Measured and (b) reconstructed spectrogram. (c) Measured (solid line) and reconstructed (dash line) spectrum. The central wavelength is around 400 nm. (d) Reconstructed temporal intensity envelope. The pulse duration is considered as 25.8±0.6 fs FWHM from fitting the main pulse to a Gaussian function.

## 2. Calibration of Time Zero and Temporal Resolution for Time-Resolved XUV Absorption Measurements

The time zero and temporal resolution (instrument response function) is calibrated with a cross-correlation measurement of pump and probe pulses in He 2s2p state.

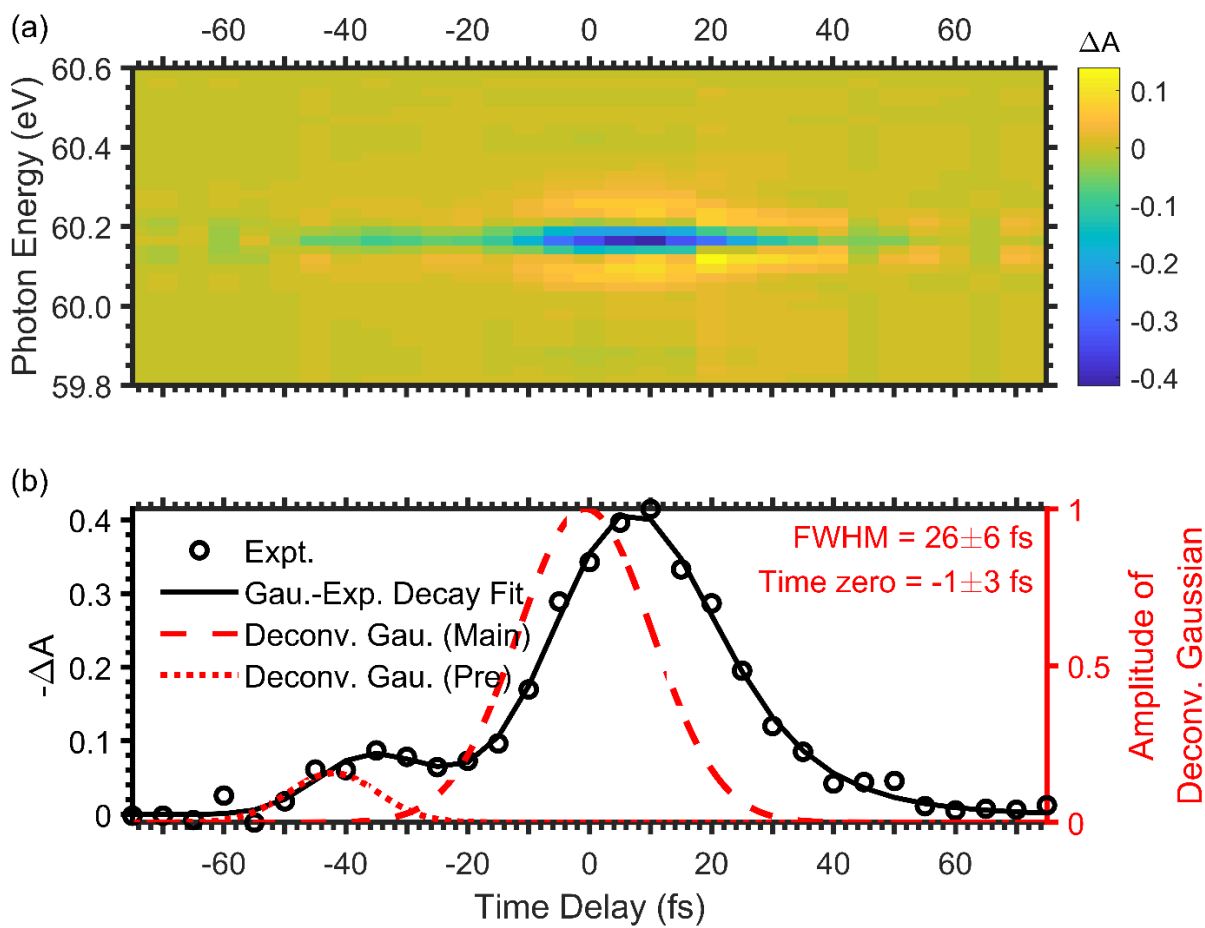

**Figure S7.** The cross-correlation measurement of pump and probe pulses in He 2s2p state. (a) Time-resolved XUV differential absorption (ΔA) spectrum of He 2s2p state. The XUV absorption is suppressed (ΔA turns negative) by the pump pulse when pump and probe pulses overlap. (b) Circle: Negative of measured differential absorption (-ΔA) at 60.16 eV. Black solid line: Fit by a convolution of two Gaussian (main and pre-pulse) and an exponential decay function in time delay. Red dash line: Deconvoluted Gaussian (main pulse). The fitted FWHM and center of this Gaussian is 26±6 fs and -1±3 fs, and these are taken as the instrument response function and time zero for this work, respectively. Red dash line: Deconvoluted Gaussian (pre-pulse)

## 3. Calibration of Spectral Axis

In this work, the XUV beam is dispersed by a grating onto a XUV CCD camera. As a result, the spectral axis of the raw data is recorded in camera pixels. We measure the static XUV absorption of He 2snp (n = 2 to 5) states, and establish the calibration curve for converting pixel to wavelength and photon energy.

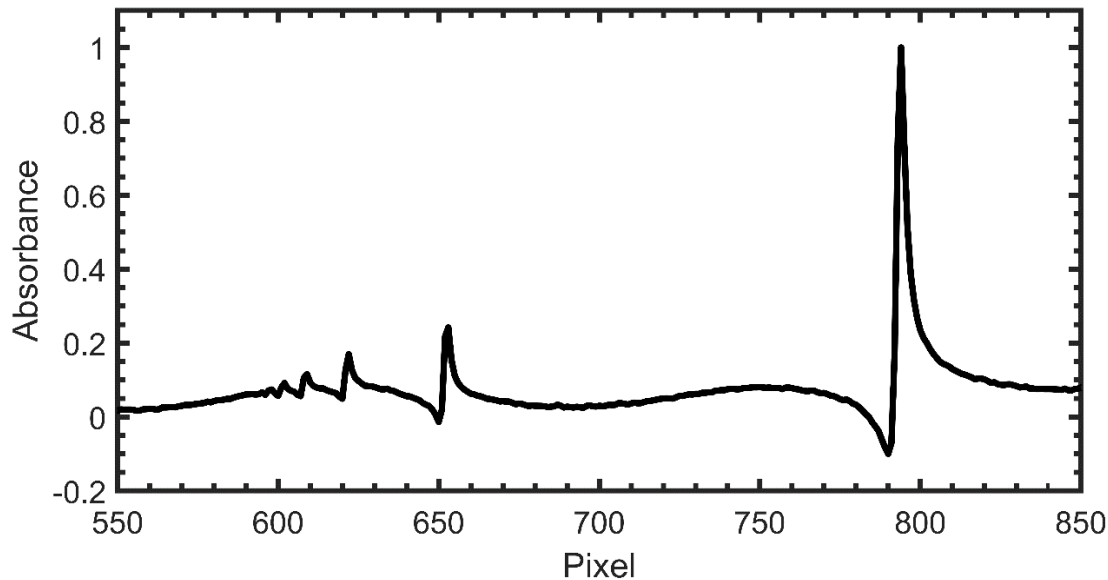

**Figure S8.** Static XUV absorption spectrum of He 2snp (n ≥ 2) excited states, as a function of XUV CCD camera pixel.

**Table S2. Calibration Table with He 2snp (n = 2 to 5) States**

| Electronic State | Resonance Location (Pixel)[a] | Resonance Energy (eV), Literature[2] | Resonance Wavelength (nm)[b] |
|:---:|:---:|:---:|:---:|
| He 2s5p | 608±1 | 64.814 | 19.129 |
| He 2s4p | 621.2±0.4 | 64.467 | 19.232 |
| He 2s3p | 652.2±0.2 | 63.658 | 19.476 |
| He 2s2p | 793.4±0.2 | 60.147 | 20.613 |

[a] Fitted to a convolution of Fano lineshape and Gaussian function, to determine the resonance location in pixel. Estimated manually for 2s5p.

[b] Converted from resonance energy (eV) in the literature.[2]

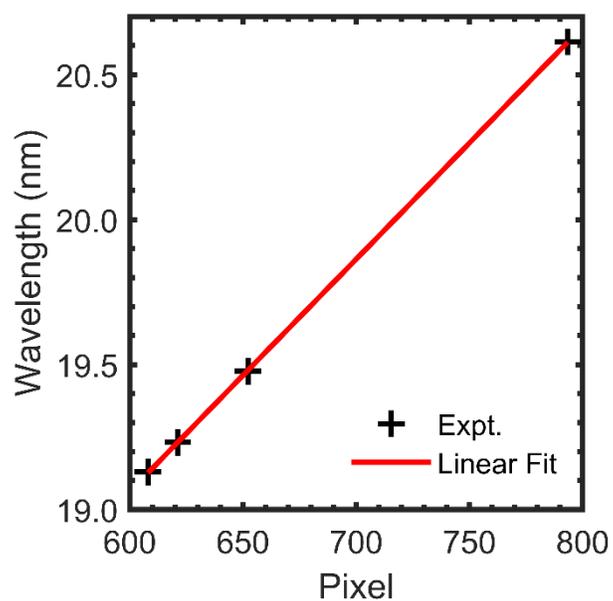

**Figure S9.** The calibration curve for wavelength vs pixel. Cross: Measured resonance location. Red line: Linear fit. This calibration curve enables the conversion from pixel to wavelength and photon energy.

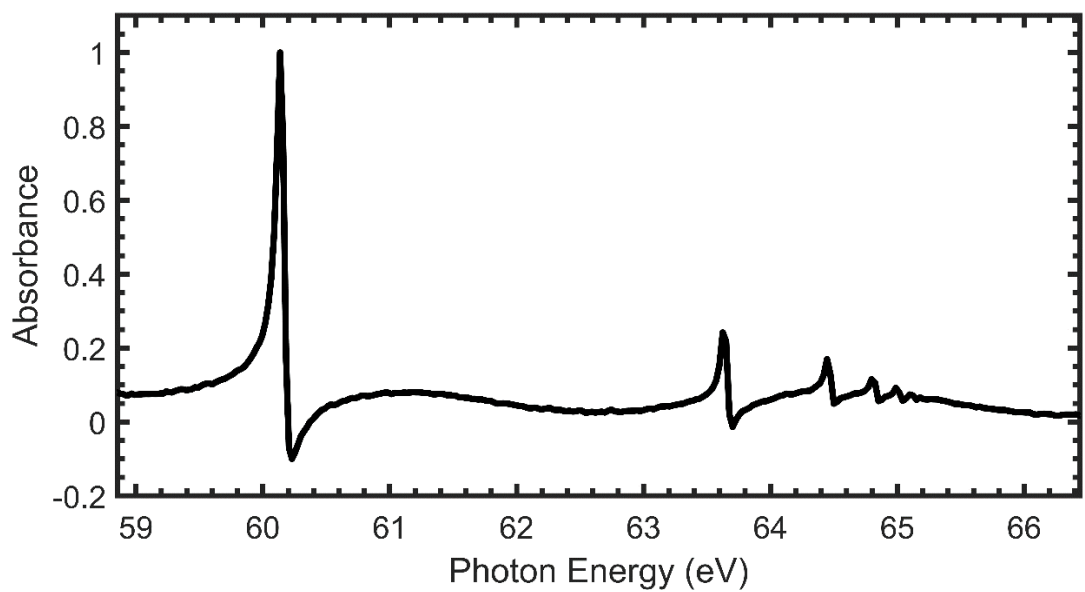

**Figure S10.** Calibrated static XUV absorption spectrum of He 2snp (n ≥ 2) excited states, as a function of photon energy.

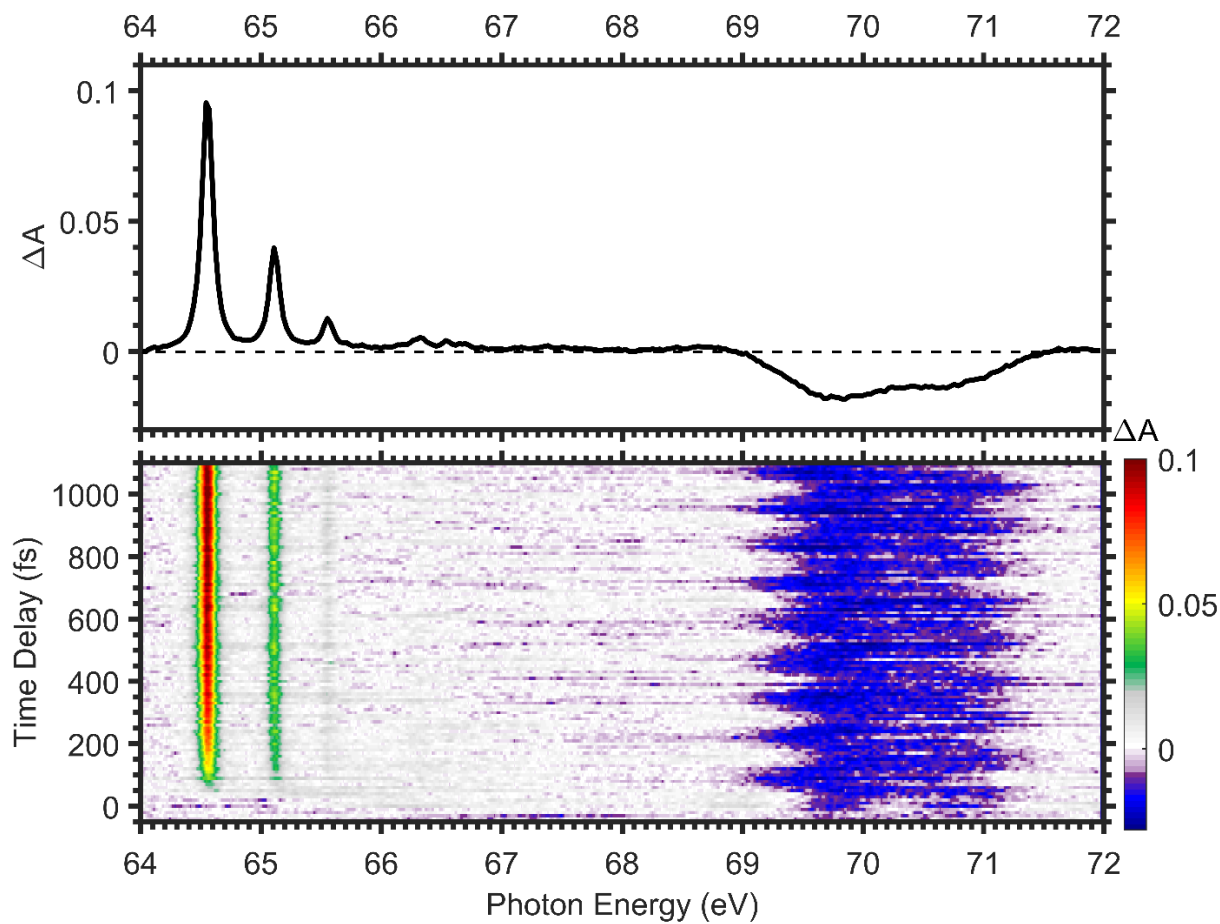

**Figure S11.** Time-resolved XUV differential absorption spectrum of CBr$_4$ with pump peak intensity 20.2×10$^{12}$ W/cm$^2$. The negative oscillating signals around 69 to 71 eV encode the coherent vibrational dynamics on the electronic ground state. (Upper) Differential absorption integrated from time delay of 300 to 1100 fs. Three major peaks between 64 to 66 eV correspond to dissociated atomic Br. These are likely to come from multiphoton absorption of the 400 nm pump pulse (photon energy 3.1 eV), because the lowest transition to valence-excited states (A bands) of CBr$_4$ starts at 4.9 eV (39550 cm$^{-1}$)[3,4]

**Table S3.** Energy of dissociated atomic Br 3d absorption. Our values agree reasonably with the literature. This also confirms the validity of our spectral axis calibration.

| Transition | This work Energy (eV) | Previous Literature[5] Energy (eV) |
|---|---|---|
| Br $^2P_{3/2} \rightarrow {}^2D_{3/2}$ | 65.55 | 65.58 |
| Br $^2P_{1/2} \rightarrow {}^2D_{3/2}$ | 65.11 | 65.13 |
| Br $^2P_{3/2} \rightarrow {}^2D_{5/2}$ | 64.56 | 64.54 |

**4. Measured Relative Slope of Br-3d Core-Excited State Potential Energy Surface (PES)**

**Table S4.** Measured relative slope of (Br-3d$_{3/2,5/2}$)$^{-1}$10a$_1^*$ core-excited state PES along the symmetric stretch mode. Relative slope is determined experimentally as the ratio of measured oscillation amplitude of XUV absorption energy $\Delta E_{\text{XUV}}$ (Br-3d$_{5/2}$)/ $\Delta E_{\text{XUV}}$ (Br-3d$_{3/2}$).

| Scan # | Peak Intensity ($\times 10^{12}$ W/cm²) | $\Delta E_{\text{XUV}}$ (Br-3d$_{3/2}$) (meV) | $\Delta E_{\text{XUV}}$ (Br-3d$_{5/2}$) (meV) | Relative Slope |
|---|---|---|---|---|
| 2 | 8.15 | 2.8±0.6 | 4.1±0.9 | 1.5±0.4 |
| 3 | 10.3 | 9.6±1.2 | 9.0±1.3 | 0.9±0.2 |
| 5 | 12.5 | 7.2±1.7 | 7.6±1.6 | 1.0±0.3 |
| 4 | 19.2 | 11.9±1.9 | 12.2±1.7 | 1.0±0.2 |
| 1 | 20.2 | 10.2±1.4 | 11.3±1.4 | 1.1±0.2 |
| Mean±Error | | | | **1.1±0.13** |

## 5. Comment about Initial Phase of Wavepacket Trajectory

If the launching mechanism of the vibrational wavepacket is a pure non-resonant ISRS from a Gaussian pump pulse centered at t = 0, the expectation value of the wavepacket displacement will be a sine function with a zero initial phase.[6,7] Nuclear TDSE simulations reproduce this expectation by retrieving around 0.01 $\pi$ initial phase of wavepacket position for all intensities, with the experimental pump pulse intensity profile measured by SD-FROG (as shown in Figure S6(d)). Experimentally, over the range of intensities measured, the initial phase of XUV absorption energy varies from -0.1 to -0.6 $\pi$ (Figure S12), suggesting a possible deviation from the pure non-resonant ISRS. Potential cause may be the pre-pulse (Figure S6) or the multiphoton absorption observed in the time-resolved XUV differential absorption spectrum (Figure S11).

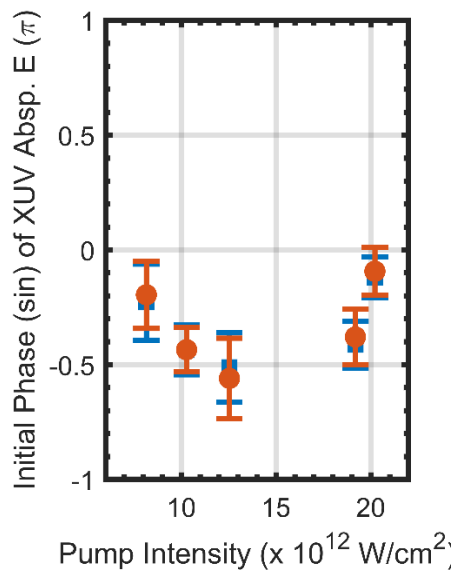

**Figure S12.** The initial phase at time zero from fitting the measured XUV absorption energy as a function of time-delay to a sine function for different pump-pulse intensities.

## 6. Nuclear Time-Dependent Schrodinger Equation Simulations

We simulated the dynamics of the symmetric stretch mode in a manner similar to what was described in a previous work.[8] In brief, under the harmonic approximation (which is valid in the small-displacement limit), the normal modes are all independent of each other and can be treated separately. The symmetric stretching mode evolves under the influence of the Hamiltonian:

$$\hat{H}(t) = \frac{\hat{p}^2}{2\mu} + \frac{1}{2}\mu\omega_0^2\,\hat{q}^2 - \frac{1}{2}\alpha(\hat{q})|\mathcal{E}(t)|^2$$

where $\hat{p}$ is the momentum operator along this mode, $\hat{q}$ is the operator for bond stretching (relative to the equilibrium bond length), $\alpha(\hat{q})$ is the isotropic polarizability as a function of $\hat{q}$, $\mu$ is the normal mode mass, $\omega_0$ is the normal mode frequency, and $|\mathcal{E}(t)|^2 = I(t)$ is the temporal intensity envelope of the pump pulse. Details of the above terms are as follows.

The isotropic polarizability $\alpha(\hat{q})$ tends to increase with bond stretching as the bonding electron density becomes "softer" (i.e. more diffuse) in response.[9] For small displacements, it can be approximated with $\alpha(q) = \alpha_0 + \alpha_1 q + \alpha_2 q^2$. Fitting SCAN0/decontracted aug-cc-pVTZ polarizabilities across 201 uniformly spaced bond stretches spanning ±0.1 Å relative to equilibrium for CBr$_4$ yields $\alpha_0 = 97.76, \alpha_1 = 38.70, \alpha_2 = 16.90$, in atomic units.

The normal mode mass $\mu$ = 575437.04 atomic units (i. e. 4 times the mass of a single $^{79}$Br atom, assuming all four Br atoms in the molecule correspond to this isotope). The normal mode frequency is 0.00129 atomic units (282.32 cm$^{-1}$) from the ground state DFT calculations.

The temporal intensity envelope of the pump pulse $|\mathcal{E}(t)|^2 = I(t)$ is calculated as

$$I(t) = I_{pk}(\text{a.u.})I_{FROG}(t)\cos^2(\omega t)$$

where $I_{pk}$ (a.u.) is the calibrated peak intensity in atomic units (Table S1), $I_{FROG}(t)$ is the normalized temporal intensity envelope from SD-FROG measurement (Figure S6(d)), $\omega$ is the angular frequency of 400 nm pump pulse and equals 0.1139 atomic units.

With these parameters, the time-dependent Schrödinger equation (TDSE) for the symmetric stretching mode was numerically simulated within the interaction picture for each of the experimentally studied pump laser intensities, taking the field-free harmonic oscillator terms as the reference Hamiltonian and the $-\frac{1}{2}\alpha(\hat{q})|\mathcal{E}(t)|^2$ term as the perturbation. The simulations were done in the basis of the ten lowest energy eigenstates of the unperturbed harmonic oscillator Hamiltonian. The wavefunction at t = −4134.14 atomic units (100 fs) was assumed to be the ground state of the unperturbed harmonic oscillator, and the wave-function was subsequently propagated in time to t = 41339.86 atomic units (1000 fs) with step sizes of 2 atomic units (0.048 fs).

## 7. Comment about Polarizability Calculations

The TDSE simulations along the symmetric stretch mode appear to predict bond stretches that are roughly three times as great for a given pump power intensity as those estimated from the experimental $\Delta E_{XUV}$ values and the computed Br core-excited PES slope of -9.4 eV/Å. A potential source of this discrepancy might be the use of *too large* polarizabilities for the TDSE simulations, as more simplified previous models[7] have indicated that the bond stretch amplitude is directly proportional to $\alpha_1$ (i.e. the slope of the polarizability vs bond stretch at the equilibrium geometry).

Such an overestimation could potentially arise from either errors in the finite difference approximation used for computing polarizabilities or errors intrinsic to the SCAN0/decontracted aug-cc-pVTZ model chemistry. Both scenarios appear to be unlikely, as discussed below.

**7.1 Finite difference:** Q-Chem presently lacks analytic polarizabilities with the X2C relativistic method. As a result, the polarizability was computed via finite differences of the energy with different electric fields. As the polarizability is the negative of the second derivative of the energy vs applied field strength, we can utilize standard finite difference formulae and obtain:

$$\alpha_{zz} = -\left(\frac{d^2}{d\mathcal{E}_z^2}E(\mathcal{E}_z)\right)_{\mathcal{E}_z=0} = \frac{30E(0) - 16E(h) - 16E(-h) + E(2h) + E(-2h)}{12h^2} + O(h^4)$$

where $E(\mathcal{E}_z)$ is the energy of the system computed with an applied field $\mathcal{E}_z$ along the z direction and $h = 0.005$ atomic units is the applied electric field strength for the finite difference calculations. As CBr4 is tetrahedral, the isotropic polarizability $\alpha = \alpha_{zz}$. Furthermore, the lack of a permanent dipole moment ensures $E(h) = E(-h)$. So the previous five-point stencil simplifies to the following three point formula:

$$\alpha_{zz} = -\left(\frac{d^2}{d\mathcal{E}_z^2}E(\mathcal{E}_z)\right)_{\mathcal{E}_z=0} = \frac{15E(0) - 16E(h) + E(2h)}{6h^2} + O(h^4)$$

As a result, for any given bond stretch, the energies are computed for zero external fields, and with fields of 0.005 a.u. and 0.01 a.u. along the z direction ($E(0), E(h)$ and $E(2h)$, respectively) to compute the polarizability. The scale of the finite difference error can be estimated by comparing the finite difference results in the nonrelativistic (i.e. X2C free) SCAN0/decontracted aug-cc-pVTZ case, for which analytic values can be computed. The resulting values for the 201 uniformly spaced bond stretches spanning ±0.1 Å relative to the equilibrium geometry that were used to fit the quadratic polynomial $\alpha(q) = \alpha_0 + \alpha_1 q + \alpha_2 q^2$ indicate very little finite difference error. Specifically, the largest deviation between analytic and finite difference nonrelativistic SCAN0/decontracted aug-cc-pVTZ isotropic polarizabilities is 0.0008 atomic units for the geometries utilized and fits from the two approaches predict identical $\alpha_1$ values of 38.18 atomic units. Finite difference errors in the X2C SCAN0/decontracted aug-cc-pVTZ isotropic polarizabilities therefore appear to be not a major concern. We also take the opportunity to note that relativistic effects appear to slightly increase the isotropic polarizability in this molecule. All the computed polarizability values are provided in the accompanying Excel spreadsheet.

**7.2 Electronic Structure:** The other potential source for errors in the polarizability may be inadequacies of the SCAN0/decontracted aug-cc-pVTZ model chemistry, even though it has been shown to be accurate for lighter elements[10]. We therefore carried out nonrelativistic CCSD(T)/aug-cc-pVTZ calculations (utilizing the aforementioned finite difference formula) for a few bond stretches to determine if SCAN0/decontracted aug-cc-pVTZ was acceptable. Using 41 equally spaced geometries between 1.9046-1.946 Å (same as those used in the main text for finding the Br core-excited PES slope from ROKS), we find that nonrelativistic CCSD(T)/aug-cc-pVTZ yields very similar polarizabilities (albeit slightly smaller by 0.1-0.2 atomic units) as nonrelativistic SCAN0/decontracted aug-cc-pVTZ. The fitted $\alpha_1$ values in this interval are: 38.02 atomic units from SCAN0/decontracted aug-cc-pVTZ and 36.64 atomic units from CCSD(T)/aug-cc-pVTZ. However, the CCSD(T) calculations utilize a frozen core (C 1s, Br 1s 2s 2p) and the decontracted aug-cc-pVTZ basis used in the SCAN0 calculations is more flexible (and hence likely to lead to larger polarizabilities) than the aug-cc-pVTZ basis used for the CCSD(T) calculations. It nonetheless appears clear that the polarizabilities obtained from finite differences on SCAN0/decontracted aug-cc-pVTZ with X2C are quite reasonable, and lead to $\alpha_1$ within 5% of CCSD(T). The polarizability fits are therefore extremely unlikely to contribute to the factor of 3 difference between the bond stretching amplitude predicted by TDSE simulations and experiment.